\documentclass{bmcart}
\usepackage{mathtools}
\usepackage[utf8]{inputenc} 

\usepackage{amsmath}
\usepackage{ctable}
\usepackage{url}

\begin{document}

\begin{frontmatter}

\begin{fmbox}
\dochead{Research}


\title{Large-scale and high-resolution analysis of food purchases and health outcomes}


\author[
   addressref={aff1},
	 corref={aff1},
   email={lajello@gmail.com}
]{\inits{LMA}\fnm{Luca Maria} \snm{Aiello}}
\author[
   addressref={aff2},
   email={schifane@di.unito.it}
]{\inits{RS}\fnm{Rossano} \snm{Schifanella}}
\author[
   addressref={aff1},
   email={quercia@cantab.net}
]{\inits{DQ}\fnm{Daniele} \snm{Quercia}}
\author[
   addressref={aff3},
   email={lucia.delprete@tesco.com}
]{\inits{LDP}\fnm{Lucia} \snm{Del Prete}}


\address[id=aff1]{%
  \orgname{Nokia Bell Labs},
  \city{London},
  \cny{UK}
}
\address[id=aff2]{%
  \orgname{University of Turin},
  \city{Turin},
  \cny{Italy}
}
\address[id=aff3]{
  \orgname{Tesco Labs}, 
  \city{London},                              
  \cny{UK}                                    
}



\end{fmbox}


\begin{abstractbox}


\begin{abstract}
To complement traditional dietary surveys, which are costly and of limited scale, researchers have resorted to digital data to infer the impact of eating habits on people's health. However, online studies are limited in resolution: they are carried out at country or regional level and do not capture precisely the composition of the food consumed. We study the association between food consumption (derived from the loyalty cards of the main grocery retailer in London) and health outcomes (derived from publicly-available medical prescription records of all general practitioners in the city). The scale and granularity of our analysis is unprecedented: we analyze 1.6B food item purchases and 1.1B medical prescriptions for the entire city of London over the course of one year. By studying food consumption down to the level of nutrients, we show that nutrient diversity and amount of calories are the two strongest predictors of the prevalence of  three diseases related to what is called the ``metabolic syndrome'': hypertension, high cholesterol, and diabetes. This syndrome is a cluster of symptoms generally associated with obesity, is common across the rich world, and affects one in four adults in the UK. Our linear regression models achieve an $R^2$ of 0.6 when estimating the prevalence of diabetes in nearly 1000 census areas in London, and a classifier can identify (un)healthy areas  with up to 91\% accuracy. Interestingly, healthy areas are not necessarily well-off (income matters less than what one would expect) and have distinctive features: they tend to systematically eat less carbohydrates and sugar, diversify nutrients, and avoid large quantities. More generally, our study shows that analytics of digital records of grocery purchases can be used as a cheap and scalable tool for health surveillance and, upon these records,  different stakeholders from governments to insurance companies to food companies could implement effective prevention strategies.
\end{abstract}




\end{abstractbox}

\end{frontmatter}

\section{Introduction}
More than 300k premature deaths in Europe are caused by obesity~\cite{ekelund2015physical}. In the United Sates, 36\% of adults and 17\% of children are not just overweight but obese~\cite{Flegal16}. In UK one in four adults is obese~\cite{van2014prevalence} and it is estimated that more than half of European citizens will be obese by 2050. Obesity has long term costs. It raises the risks of diabetes and heart diseases, which result in increased health-care spending (70B of Euros in Europe every year), and ultimately cost lives.

Healthy eating is one of the most effective intervention to counter such risks~\cite{kaur2014comprehensive}. In the developing world, people can now afford to eat more food, particularly processed food high in fat and sugar. Monitoring dietary habits of people and persuading them to eat better and exercise more ranks high on the lists of priorities for governments around the world. 

Factors associated with food-related disorders are hard to untangle. On top of that, many studies about dietary habits rely on data limited in scale. To partly address the lack of data, computer scientists have recently resorted to the Web. They have analyzed nutrition sites containing food recipes across the world~\cite{trattner2017investigating}, and food images posted on social media~\cite{dechoundury2016characterizing,ofli2017saki}, and they have done so to infer what Web users are likely to eat. Approaches based on this type of data either suffer  from limited spatial resolution (they provide reliable estimates of food consumption but do so at  a geographic resolution no lower than city level) or, given their biases, cannot capture reliably what actually people eat. 

To further those studies, we explore fine-grained associations between food purchases and disease prevalence at the level of Middle Super Output Areas (MSOA) for the entire city of London. To do that, we analyze, for the first time, the purchases derived from the  loyalty cards of the main grocery retailer in the country and  match them with the prevalence of the three main medical conditions associated with what is called  the ``metabolic syndrome''.  ``Syndrome'' is the medical term for a collection of symptoms whose common cause is not properly understood, and the ``metabolic syndrome'' is  a cluster of (obesity) symptoms that includes hypertension, cholesterol, and diabetes. This syndrome is common in rich countries and affects one in four adults in the UK. The prevalence of these symptoms is derived from  prescription data made publicly available by all the general practitioners in the city. In so doing, we make four main contributions:

\begin{itemize}

\item Based on the literature, we formulate seven main research questions that relate food consumption to the three diseases, and operationalize six metrics to answer them (\S\ref{sec:questions}).

\item We combine two sets of geo-referenced data in London (\S\ref{sec:methods}). One set contains records of every single food item customers in bought in 2015 at the largest grocery retailer in the country and the second set contains every single medical prescription written in 2016 by all London's General Practitioners (medical doctors or, simply, GPs)\footnote{We also conducted all the analyses using prescription data from 2015, and obtained similar results.}. From the anonymized and aggregated food purchased data, we extract food nutrients. From the publicly available prescription data, we infer the prevalence of the three ailments: hypertension, cholesterol, and diabetes. All this data will be made available on the project's site \footnote{\url{http://goodcitylife.org/food}}.

\item We find that, on  average, Londoners' diet does not meet the official recommendations of the World Health Organization (\S\ref{sec:results}), in that, fat and sugar are more prevalent than what the health organization's guidelines would recommend. We also learn that  the prevalence of the three medical conditions is related  to two features, that is, it is related to food quantities (overall calories) and inversely related to nutrient diversity. These features are not only descriptive but also predictive: the $R^2$ of a linear regression that predicts diabetes in London neighborhoods reaches 0.6; also, a binary classifier can identify (un)healthy areas from their food consumption with an accuracy as high as  91\%. 

\item We conclude by showing how our methodology and findings might improve evidence-based public outreach initiatives and  inform the design of new consumer technologies (\S\ref{sec:discussion}). 
\end{itemize}

\section{Related work}
It is assumed that people are able to freely choose what they eat, yet  that does not entirely reflect reality. The amount of food consumed by an individual is also influenced by: \emph{i)} the habits and associations around food formed at a young age~\cite{boumtje2005dietary}; \emph{ii)} external factors such as portion sizes, food cost and availability~\cite{lee2012role}; and \emph{iii)} biological factors (e.g., upon the consumption of sugary food, the brain releases dopamine, a chemical that signals pleasure and is involved in drug addiction~\cite{Volkow2012}).  To produce medical evidence on which factors impact what people eat, the gold standard is represented by randomized controlled clinical trials. Such trials are typically limited in scale and may produce conflicting results (e.g., a meta-review found that most common foods are linked to both a higher and lower risk of cancer~\cite{Schoenfeld13}). 

The availability of large datasets now makes it possible to study health outcomes at unprecedented scales~\cite{ahnert2013network}. However, health records of individual patients are not widely available, not least because of privacy concerns. By contrast,  a variety of Web sites have been  proven to be a good source of data for public health surveillance:  search query logs have been used to forecast the spreading of influenza epidemics~\cite{ginsberg2009detecting}; microblogging platforms such as Twitter have been used to monitor public health at scale~\cite{prier2011identifying,paul2011you} and to estimate the prevalence of a wide range of pathologies, from mental illnesses~\cite{de2013predicting} to obesity~\cite{mejova2015foodporn}; and pictures on social media have been used to estimate  values of Body Mass Index (BMI)~\cite{weber2016crowdsourcing,kocabey2017face}. 
 
On the web, in addition to communities of general interest, there are  communities specialized in food. Such  communities  have allowed researchers to study dietary patterns~\cite{west13cookies,wagner2014nature,sajadmanesh17kissing}, and how these patterns change depending on a country's culture~\cite{ahn2011flavor}.  Databases containing millions of receipts have been used to quantify each receipt's healthiness based on its ingredients and associated images~\cite{said2014you,trattner2017investigating}, and the resulting proxies for healthiness have been recently incorporated into food recommender systems~\cite{ge15health}. 

Given its scale, web data allows for the study of various societal aspects concerning food consumption. After collecting food-related tweets in 50 US states, researchers found that caloric values of the foods mentioned in the tweets related to state-wide obesity rates~\cite{abbar2015you}. In USA, large areas suffer from obesity: food deserts, for example, are areas that have limited access to nutritious food and typically happen to be of low income. Recently, De Choudhury \emph{et al.} analyzed millions of food-related Instagram images~\cite{sharma2015measuring} and found that food deserts indeed consume food high in fat, cholesterol, and sugar more than the other locations do~\cite{dechoundury2016characterizing}. In addition to what social media users might eat, images might reveal perceptions.  Indeed, from a food image, one can infer two quantities: first, how healthy the food in the image is (based on what the image depicts); and second, how healthy the social media user perceives it to be (based on the user's comments).  Researchers have computed these two quantities for a variety of countries and found that the gap between the two relates to health outcomes~\cite{ofli2017saki}. All these works which have relied on Web data, however, invariably suffer from a number of self-presentation and self-selection biases and, as such, the resulting datasets might not reflect actual food consumption~\cite{mejova2016fetishizing,wagner15men}.

To partly address these biases, a few studies have analyzed data of grocery purchases at scale. Nevalainen \emph{et al.} collected loyalty card data in Finland voluntarily provided by more than 13K customers, and analyzed consumptions across categories~\cite{nevalainen2018large}. Guidotti \emph{et al.} mined millions of transactions recorded by two retailers to train algorithms that suggest what online shoppers might like~\cite{guidotti2018personalized}. Mamiya \emph{et al.} collected grocery purchase data from multiple stores in the metropolitan area of Montreal (Canada), performed an extensive analysis of consumption of carbonated drinks~\cite{mamiya2017novel}, and found that education inversely correlates with the consumption of soft drinks~\cite{mamiya2018susceptibility}. None of these studies, however, have related food consumption to health outcomes.

To sum up, prior work has shown that controlled dietary studies at scale are costly and, as such, are rare, and that social media data---albeit useful to better understand nutrition habits (especially across countries and cultures)---is affected by a number of biases and typically comes at coarse-grained spatial resolutions.  What is needed is a new approach to measuring at scale what real people (as opposed to study participants) eat and drink, ideally under naturalistic conditions.  Before introducing the datasets and methods with which we tackle this challenge, we introduce our seven research questions.

\section{Research questions} \label{sec:questions}
To begin with, let us look at one of the simplest food-health relationships: people consume more calories than they use, and the surplus is stored as fat~\cite{greger2015not}. We test that with our first question:

\vspace{3pt}\textbf{RQ1:} \textit{Is calorie consumption positively associated with hypertension, high cholesterol, and diabetes?}

\vspace{3pt}Yet that could only be part of the story. It might be less about the number of calories than about their concentration. When one regularly eats calorie-dense animal products and junk foods, what changes is not only the taste buds but also the brain chemistry. Calorie-diluted foods (e.g., green smoothies) do not lead to a dopamine response but calorie-dense foods (e.g., ice creams) with the same amount of calories do. Fatty and sugary foods are energy dense, and their overconsumption has often been compared to drug addiction~\cite{Volkow2012}. To put it simply, given two foods with the same amount of calories but different concentrations, the delivery of pleasure within people's brains is quicker for the calorie-dense food. So our second research question is: 

\vspace{3pt}\textbf{RQ2:} \textit{Is calorie concentration positively associated with the three medical conditions?}

\vspace{3pt}Not all calories are created equal though. The U.S. government's official Dietary Guidelines for Americans recommends the reduction of sugar, calories, saturated fat, sodium, and trans fat and, at the same time, it recommends increasing fibres, of which at least ``a quarter of the American population is not reaching an adequate intake''~\cite{greger2015not}. Therefore, it would make sense to look at individual nutrients, and we do so next.  The digestive system breaks down \emph{carbohydrates} into a simple sugar called glucose. To get from the bloodstream into your cells, glucose requires insulin. Without insulin, the glucose builds up in the blood. Inappropriate \emph{fat} storage may keep cells from responding properly to insulin, causing insulin resistance. Eventually blood-sugar levels rise out of control and the patient develops diabetes. Even among healthy individuals, a high-fat diet impairs the body's ability to handle \emph{sugar}. But it is not all to do with fat. Obesity might be caused by diets rich in carbohydrates and sugar in the first place. Interestingly, sugar seems to change the brain's circuitry~\cite{avena08,iozzo12}. When people consume a sugary food, their brains release dopamine~\cite{iozzo12}, which signals pleasure. This evidence leads to our third research question:

\vspace{3pt}\textbf{RQ3:} \textit{Are fat, carbohydrates and sugar positively associated with the three medical conditions?}

\vspace{3pt}Not all fats affect the muscle cells in the same way. Palmitate (the saturated fat found mostly in meat, diary, and eggs) causes insulin resistance. On the other hand, oleate (the mono-unsaturated fat found mostly in nuts, olives, and avocados) protects us against the detrimental effects of saturated fat. Research findings on the impact of saturated fats are controversial though. In 2014, a large meta-analysis showed no relationship between saturated fats and heart disease~\cite{Chowdhury14}. Hence:

\vspace{3pt}\textbf{RQ4:} \textit{What is the relationship between saturated fats and the three medical conditions?}

\vspace{3pt}On a more positive note, consider fibres. Humans evolved over millions of years eating mostly wild plants, likely in excess of one hundred grams daily~\cite{clemens12}. That is much more than what the average person eats today. Given their health benefits, we posit the following question:

\vspace{3pt}\textbf{RQ5:} \textit{Are fibres negatively associated with the three medical conditions?}

\vspace{3pt}Going beyond individual nutrients, one could study their composite impact. More specifically, research has shown that healthy diets is associated with diversity of foods~\cite{ashima93}. Therefore, our next research question is:

\vspace{3pt}\textbf{RQ6:} \textit{Is nutrient diversity negatively associated with the three medical conditions?}

Finally, the amount of food people consume is often influenced by external factors, including the size of their plate~\cite{Wansink06}. Our final research question is then:

\vspace{3pt}\textbf{RQ7:} \textit{Is the overall weight of food consumed positively associated with the three medical conditions?}

\section{Methods}\label{sec:methods}

\subsection{Datasets}\label{sec:methods:datasets}

\subsubsection{Food purchases}\label{sec:methods:datasets:food}

At all our retailer's 411 shops in Greater London, 1.6M customers used their loyalty cards and bought 1.6B food products in the entire year of 2015. Given the use of loyalty cards, purchase data is stored in the following anonymized form: customer postcode area/region, store postcode, productID, and timestamp. Each productID is associated with the product's net weight, total energy, fats, saturated fats, carbohydrates, sugars, proteins, and fibres. The last six elements are expressed as grams of substances contained in the product. Using standard guidelines~\cite{whitney2007understanding}, we map grams into corresponding calories by simply multiplying them by fixed factors: 9 Kcals per gram for fats, 4 Kcals for proteins and carbohydrates, and 2 Kcals for fibres\footnote{Fibres have a calorie intake of either 2 or 0 Kcals depending on the type of fibre, which is quite small since they mostly go through the digestive system without being assimilated.}.

\subsubsection{Chronic diseases}\label{sec:methods:datasets:diseases}

At all the 1,174 general practices (GPs) in Greater London, 1.1B medicine items were prescribed in the entire 2016. Such prescription data  has been recently made publicly available\footnote{GP practice prescribing data - Presentation level: \url{https://data.gov.uk/dataset/prescribing-by-gp-practice-presentation-level}} in the following form: GP identifier, medicineID, and timestamp. Each medicineID is associated with the medicine's active ingredients, from which the corresponding diseases can be inferred. To do so, we map the prescriptions to their medicines' active ingredients and, in turn, to the chronic diseases they are supposed to treat. The mapping of an active ingredient to the most likely disease is done based on the OpenPrescribing taxonomy~\cite{curtis2018openprescribing}. As a result, for each GP, we know the number of prescriptions that are meant to treat a given disease.

We are interested not in all prescriptions but only in those related to three main factors that are generally grouped under the heading of ``metabolic syndrome'': high blood pressure (hypertension), an excess of cholesterol in the blood, and high blood-sugar levels (diabetes).

\vspace{4pt}\textbf{Hypertension}. Hypertension is a long-term medical condition in which the blood pressure in the arteries is persistently elevated. It has been identified as the most important risk factor for death in the Western world~\cite{gbd13}. To capture the prevalence of hypertension, we consider the prescriptions in three categories of the OpenPrescribing taxonomy: antihypertensive drugs (e.g., Hydralazine Hydrochloride), alpha-adrenoceptor blocking drugs (e.g., Doxazosin Mesilate), and renin-angiotensin system drugs (e.g., Lisinopril)

\vspace{4pt}\textbf{Cholesterol}. One important risk factor for coronary heart diseases is cholesterol. If cholesterol level is low, an obese and diabetic still does not develop atherosclerosis~\cite{william10,esselstyn00}. Cholesterol also seems to help some cancers migrate and invade more tissue~\cite{christiane12}. To capture the prevalence of high cholesterol, we consider one category in the OpenPrescribing taxonomy: lipid-regulating drugs (e.g., Statins).

\vspace{4pt}\textbf{Diabetes}. Diabetes is characterized by chronically elevated levels of sugar in the blood~\cite{neeland12}. Insulin is the hormone that keeps the blood sugar in check. The disease is caused by either the pancreas gland not making enough insulin (type 1 diabetes) or by the body becoming resistant to insulin's effects (type 2 diabetes, which accounts for 90-95 percent of diabetes cases~\cite{centers2017national}). Type 2 diabetes is a consequence of dietary choices (of ``high-fat and high-calorie diets'') and, as such, is preventable and often treatable. Diabetics are more likely to suffer from strokes and heart failure~\cite{pratley13}. To capture the prevalence of diabetes, we consider the prescriptions in four categories of the OpenPrescribing taxonomy: insulin, antidiabetic drugs (e.g., gliclazide), active ingredients for the treatment of hypoglycaemia (e.g., glucagon), and agents for diabetic diagnostic and monitoring (e.g., glucose blood testing reagents)\footnote{Some people using glucose testing equipment will have type 1 diabetes which is not obesity nor diet-related, but they account for less than 10\% of the cases.}.

\subsubsection{Mapping food purchases and chronic diseases}

To map all our data, we use the postcode areas/regions as our initial unit of geographic aggregation. We map our food purchases using the customers' area of residence, which are available in our grocery dataset. We then map our prescriptions based on what the anonymized prescription dataset offers - that is, based on the fraction of each GP's patients living across areas. More specifically, for each GP, we perform two steps. First, we consider the fraction of the GP's patients who live at each area since patient counts are publicly available\footnote{Numbers of Patients Registered at a GP Practice January 2015: \url{http://digital.nhs.uk/catalogue/PUB16357}} (e.g., 50\% of the GP's patients live in area $X$). Second, we assign the GP's prescriptions to an area, and the assignment is proportional to the fraction of the GP's patients who live in the area (e.g., half of the GP's prescriptions are assigned to area $X$).  We repeat these two steps for all GPs in the city. As a result, we obtain the number of prescriptions containing each medicine at each area, and normalize that number by the population, determining the per-capita prevalence of that medicine in that area.

As we shall see shortly, to support our analyzes, we need to match our data with census data, which is not available at postcode area level though. Census data is  typically defined at four different spatial resolutions\footnote{\url{https://www.ons.gov.uk/methodology/geography/ukgeographies/censusgeography}}: Lower Super Output Area (LSOA), Medium Super Output Area (MSOA), Ward, and Local Authority (LA, or, more informally, Borough). Among these four aggregations, we opt for MSOA because, at this resolution, our aggregate metrics for food consumption and disease prevalence start to be significant, not least because they concern a sufficient number of residents: our data covers 937 MSOAs in London, which have an average of 8,250 residents. As such, from now on, we refer to MSOAs as areas or neighborhoods.

Given our three diseases, we consider prescriptions related to them and express each of their prevalence at area level in terms of the number of prescriptions for each disease per capita. For example, in the case of diabetes, for any area $a$ we have:

\begin{equation}
prevalence\text{-}diabetes@a = \frac{\text{\#prescriptions for diabetes@a}}{\text{\#residents@a}}
\label{eqn:diabetes_at_area}
\end{equation}

In a similar way, we compute $prevalence\text{-}hypertension@a$ and \break $prevalence\text{-}cholesterol@a$.

\subsubsection{Socio-economic indicators}\label{sec:methods:datasets:socio}

The prevalence of chronic diseases is not only impacted by food consumption but also mediated by socio-economic conditions. Higher-income and well-educated people may have better access to doctors, gyms, parks and healthy food. There is an inverse relationship between education levels and the likelihood of getting fat in Australia, Canada and England~\cite{sassi09}. The same applies in USA: ``obesity rates in children with college-educated parents are less than half the rates of children whose parents lack a high-school degree''~\cite{economist12caveman}. In developed countries, there is a difference between cities and suburban areas: the more affluent urbanites are usually fitter than rural residents~\cite{monteiro07}. 

To control for these factors in our study, we collected data on socio-economic conditions from the 2015 UK census and from the Index of Multiple Deprivation (IMD) 2015 that is based on a basket of measures of deprivation for small areas across England\footnote{English indices of deprivation 2015: \url{https://www.gov.uk/government/statistics/english-indices-of-deprivation-2015}}. We focus on the socio-economic factors that have been found to be associated with specific food consumption patterns and, ultimately, with chronic diseases. These factors---available at MSOA level---are average income, education level, gender distribution (\%female), and average age.

\subsection{Estimating eating habits of an area}\label{sec:methods:mapping}
To estimate the eating habits of people living in an area, we pool together all the food items purchased by its residents and look at the nutritional properties of the average item. We do so by defining six metrics below.

To capture calorie consumption, we compute the average amount of calories contained in the food items purchased in an area:

\begin{equation}
calorie\text{-}consumption@a = \frac{\sum_{p \in P_a}Kcal(p)}{|P_a|}
\label{eqn:calorie_consumption}
\end{equation}

where $P_a$ is the set of all food products purchased by residents of area $a$, $p$ is one of such products, and $Kcal(p)$ is the value of kilocalories in $p$.

To capture calorie concentration rather than simple calorie counts, we compute:

\begin{equation}
calorie\text{-}concentration@a = \frac{\sum_{p \in P_a}Kcal(p)}{\sum_{p \in P_a} weight(p)},
\label{eqn:calorie_concentration}
\end{equation}

which reflects the concentration of calories in the area's ``average'' product.

For each area, we also compute the average number  of calories given by individual nutrients in a product, on average:

\begin{equation}
nutrient\text{-}calories@a = \frac{\sum_{p \in P_a}Kcal(nutrient,p)}{|P_a|},
\label{eqn:nutrient}
\end{equation}

where $P_a$ is the set of all food products purchased at area $a$; $p$ is one of such products; $Kcal(nutrient,p)$ is the energy intake given by $nutrient$ in $p$. The nutrients we consider are: fats ($fats{-}calories@a$), saturated fats ($saturated{-}calories@a$), carbohydrates ($carbs{-}calories@a$),  sugars ($sugar{-}calories@a$), \break proteins ($proteins{-}calories@a$), and fibres ($fibres{-}calories@a$).

We also capture the diversity of nutrients consumed in the area. This is computed as the Shannon entropy of the distribution of the calories given by all the nutrients:
\begin{eqnarray}
H(nutrient)@a &  =  & -\sum_{nutrient} prob(nutrient,a) \cdot \log prob(nutrient,a) 
\label{eqn:nutrients_diversity}
\end{eqnarray}
where $prob(nutrient,a)$ can be thought as the fraction of area $a$'s total calories coming from $nutrient$, which, in turn, can be written as:

\begin{eqnarray}
f_{nutrient\text{-}calories}@a & = & \frac{nutrient\text{-}calories@a}{calorie\text{-}consumption@a}
\label{eqn:nutrients_fraction}
\end{eqnarray}
For example, $f_{fat\text{-}calories}@a$ is the  fraction of $a$'s total calories  coming from fat.

\mbox{}\\
Finally, we also compute the average item weight:

\begin{equation}
item\text{-}weight@a = \frac{\sum_{p \in P_a} weight(p)}{|P_a|}.
\label{eqn:item_weight}
\end{equation}

For the sake of reproducibility of our analysis, we publicly share our data  aggregated at MSOA level\footnote{\url{http://goodcitylife.org/food}}.

\section{Results} \label{sec:results}

\subsection{Relative abundance of nutrients} \label{sec:results:nutrients}

\begin{figure}[t]
\centering
\includegraphics[width=0.99\columnwidth]{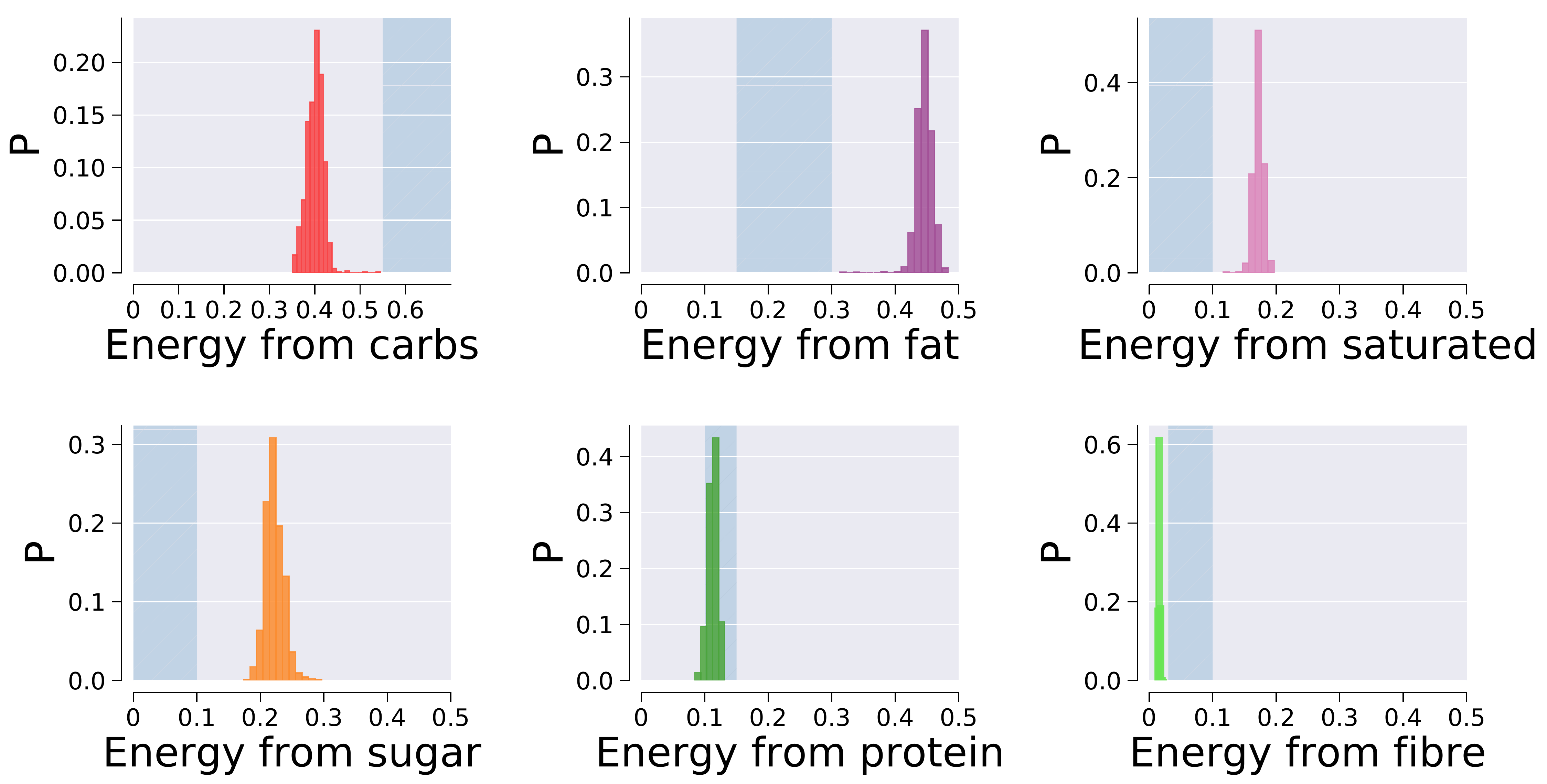}
\caption{Frequency distributions of the fraction of an area's total calories coming from each nutrient (computed with Formula~(\ref{eqn:nutrients_fraction})). The intervals recommended by the the World Health Organization are shown as dark bands.}
\label{fig:who_recs}
\end{figure}

\begin{figure}[t]
\centering
\includegraphics[width=0.45\columnwidth]{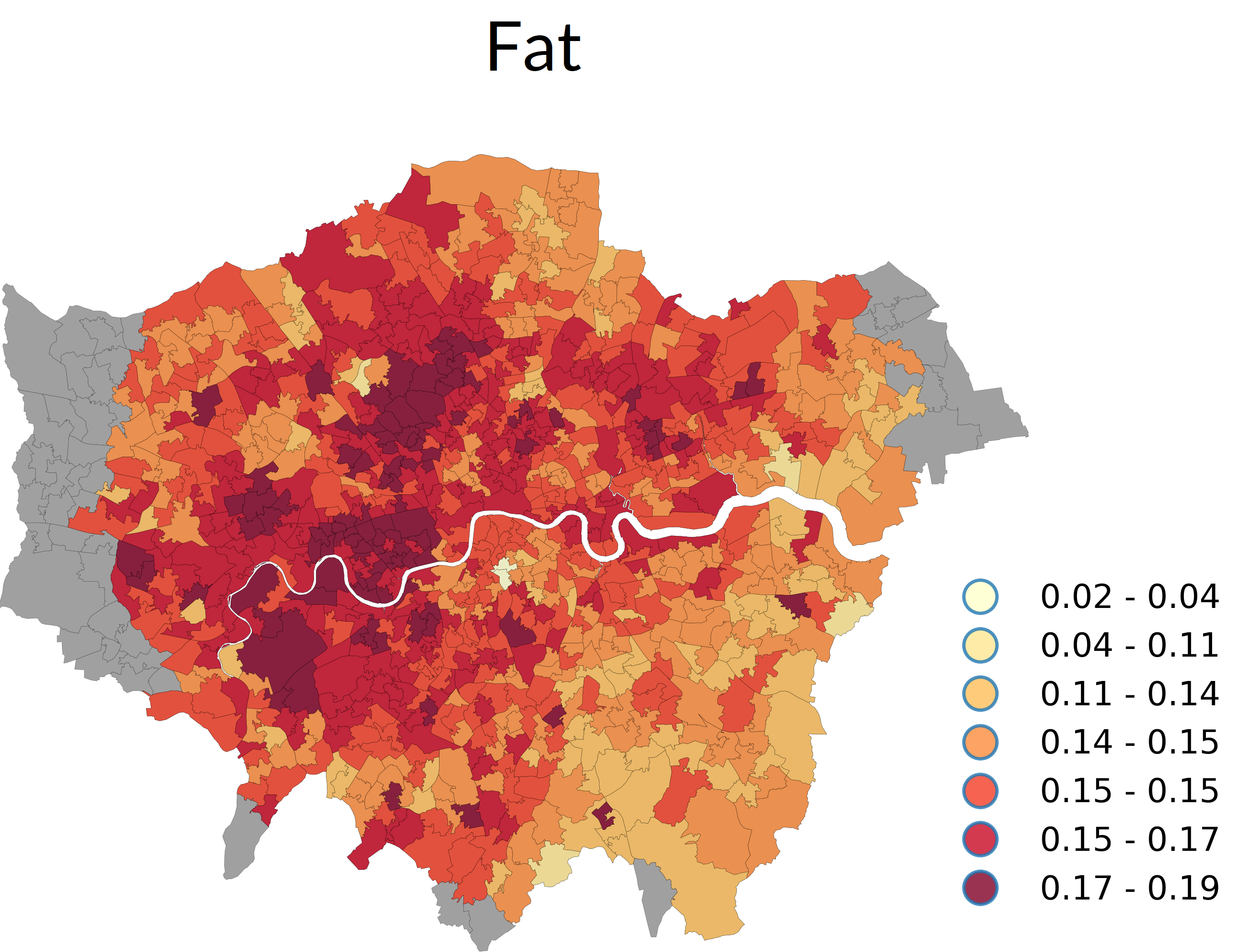}
\includegraphics[width=0.45\columnwidth]{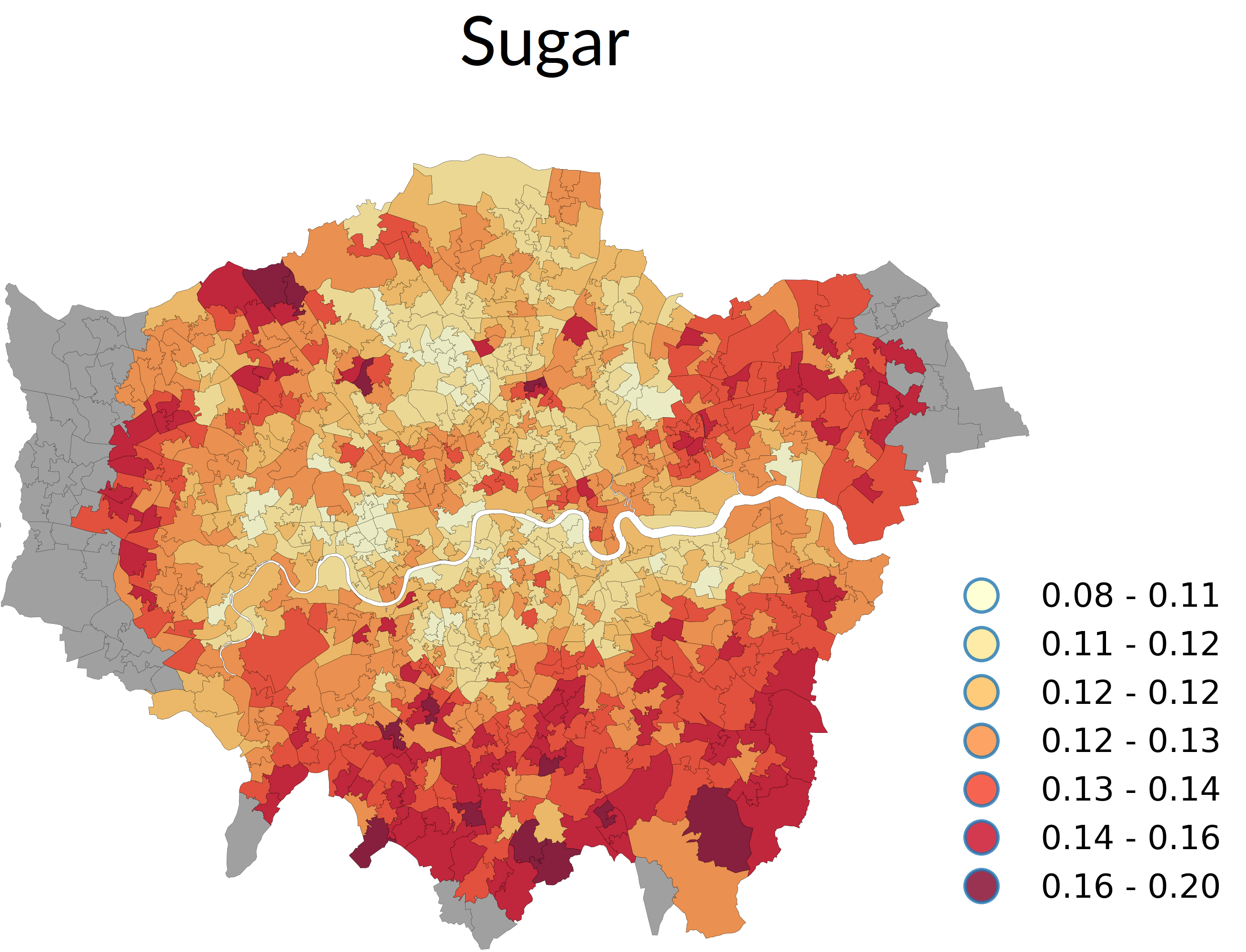}
\caption{Percentage departures from recommended limits for the consumption of fat (left) and  sugar (right). These values are computed with Formula~(\ref{eqn:departure}). Areas in red exceed the recommended limit the most (e.g., an area with 0.16 for fat is an area in which calories from fat exceed the official limit by 16\%). Areas in gray were left out because not significant.}
\label{fig:hdi_fats_sugar}
\end{figure}

The nutrition guidelines by the World Health Organization (WHO) recommend to limit the relative energy supply derived from each nutrient within specific ranges~\cite{amine2002diet}; for example, fats should contribute no more than 30\% to the total intake. By plotting the distributions of the $f_{nutrient\text{-}calories}@a$ values across neighborhoods (Figure~\ref{fig:who_recs}), one sees that Londoners, on average, buy a healthy share of protein, yet they buy unhealthy nutrients (e.g., sugar, fat, saturated fat) more than the recommended limits, and carbohydrates and fibres less than the recommended amounts. The extent to which residents collectively depart from recommended limits changes across the city  and is defined as $departure\text{-}nutrient@a$:
\begin{equation}
\begin{split}
\begin{cases} 
    | f_{nutrient\text{-}calories}@a - max_{nutrient}| \quad \phantom{\infty}\text{if}\,\, f_{nutrient\text{-}calories}@a  > max_{nutrient} \\
    0 \quad \phantom{0}  \text{if}\,\,  min_{nutrient} \leq f_{nutrient\text{-}calories}@a  \leq max_{nutrient}\\
    |  min_{nutrient} - f_{nutrient\text{-}calories}@a | \quad \phantom{0}  \text{if}\,\, f_{nutrient\text{-}calories}@a  < min_{nutrient} \\
      \end{cases}
      \label{eqn:departure}
\end{split}
\end{equation} 
That is, an area's departure from the nutrient's recommended level is zero, if the fraction of area $a$'s total calories coming from $nutrient$ is within the recommended min-max band for that nutrient (i.e., within $[min_{nutrient}, max_{nutrient}]$).  The departure from the recommended levels of, for example, fats and sugars are mapped in Figure~\ref{fig:hdi_fats_sugar}.

To  dwell on the health impact of such departure, we now match food purchases with health outcomes and start to answer our seven research questions. 

\begin{figure}[t]
\centering
\includegraphics[width=0.31\columnwidth]{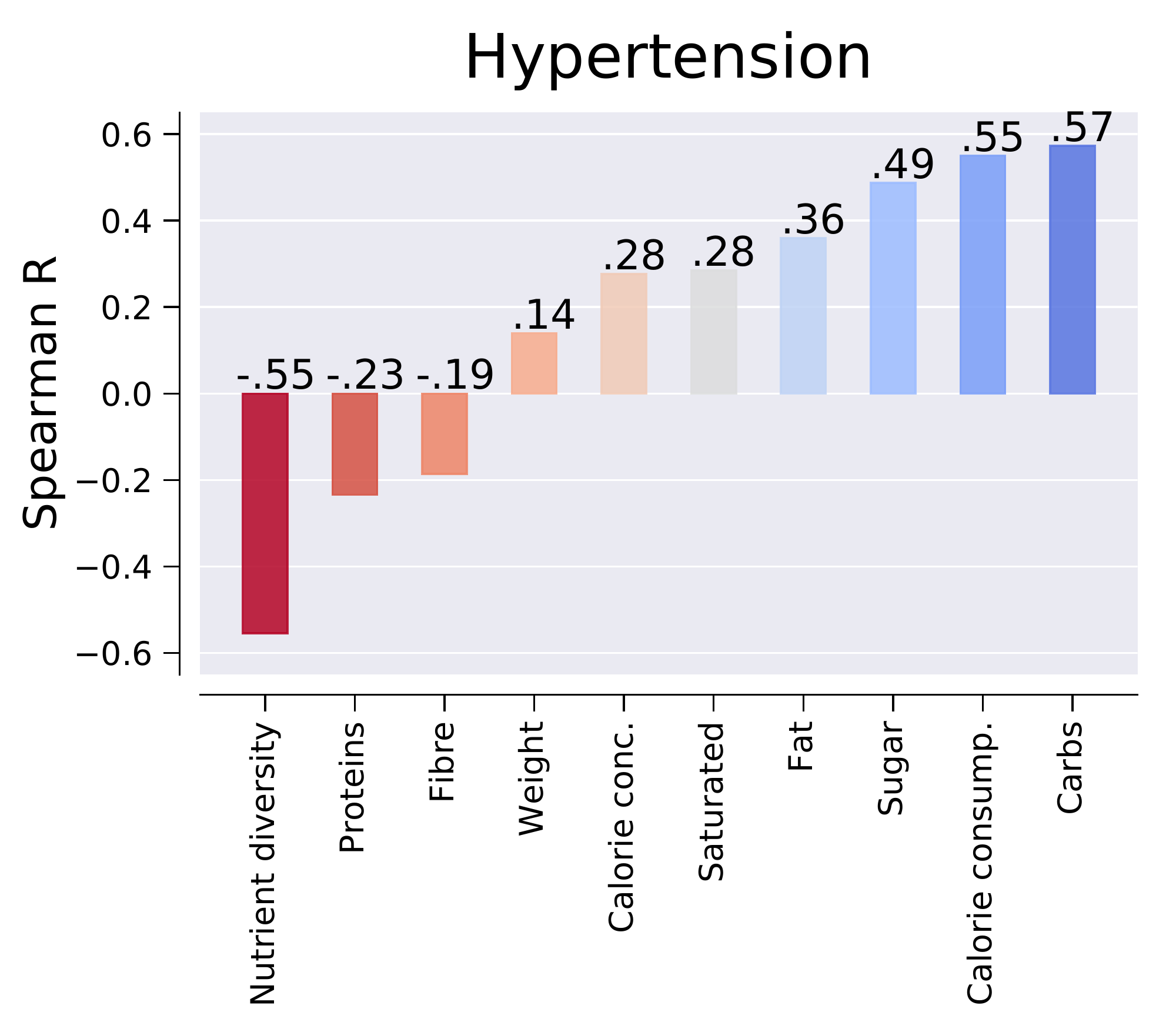}
\includegraphics[width=0.31\columnwidth]{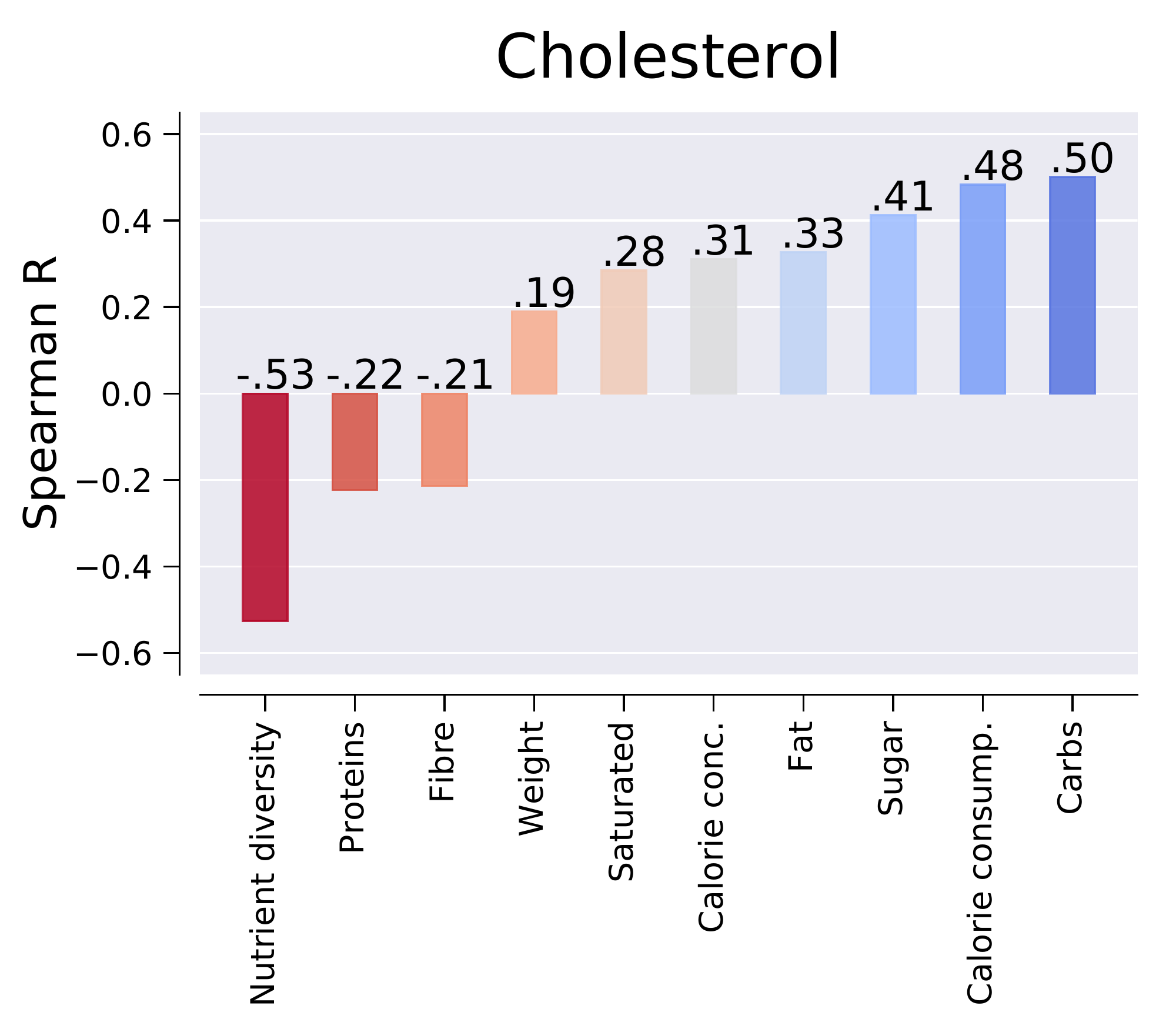}
\includegraphics[width=0.31\columnwidth]{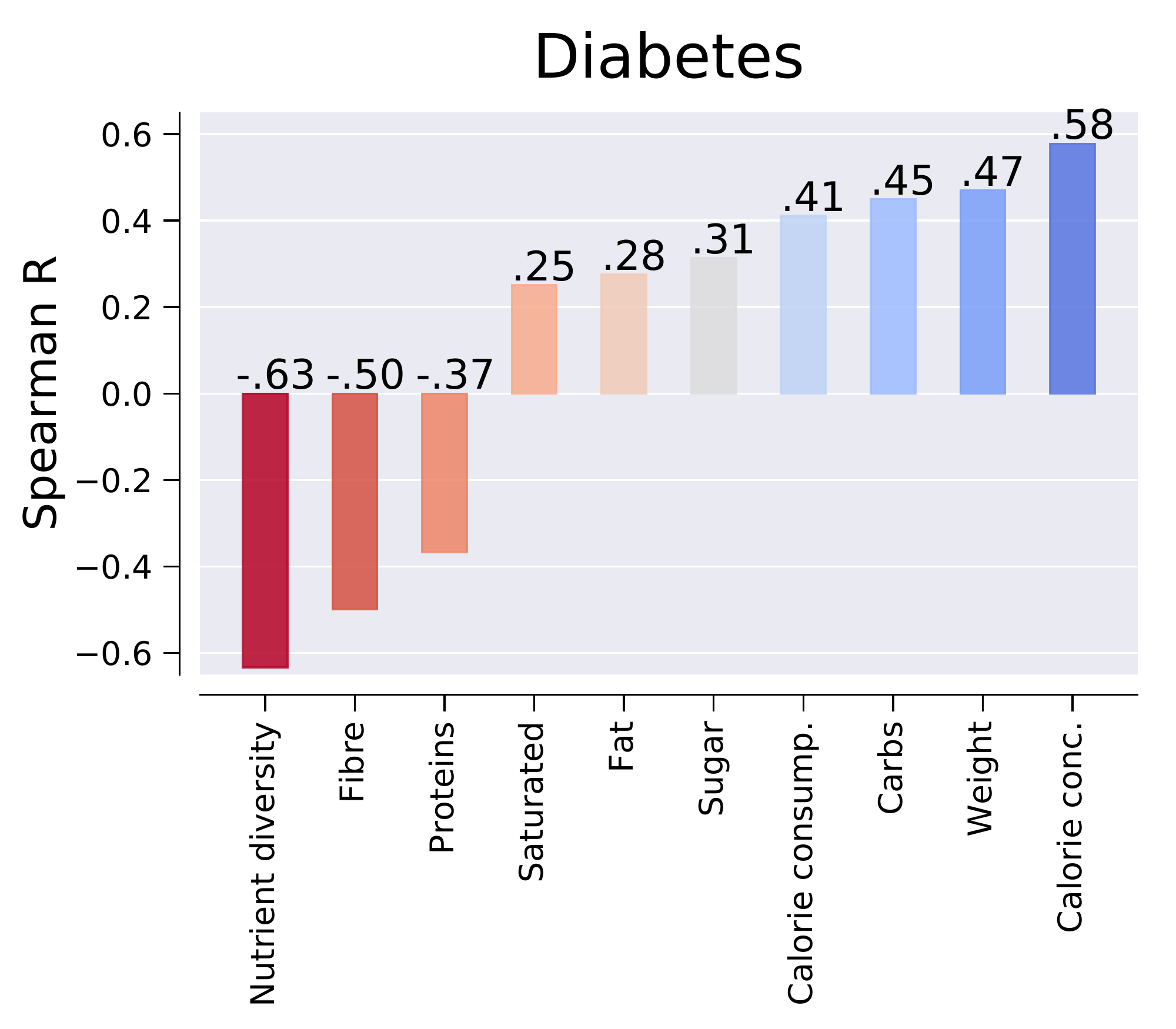}
\caption{Spearman rank correlations between disease prevalence  (as per Formula (\ref{eqn:diabetes_at_area})) and food-related metrics  (as per Formulae~(\ref{eqn:calorie_consumption})-(\ref{eqn:item_weight})). All correlations are significant with $p  < 0.001$.}
\label{fig:correlations}
\end{figure}

\begin{figure}[t!]
\centering
\includegraphics[width=0.70\columnwidth]{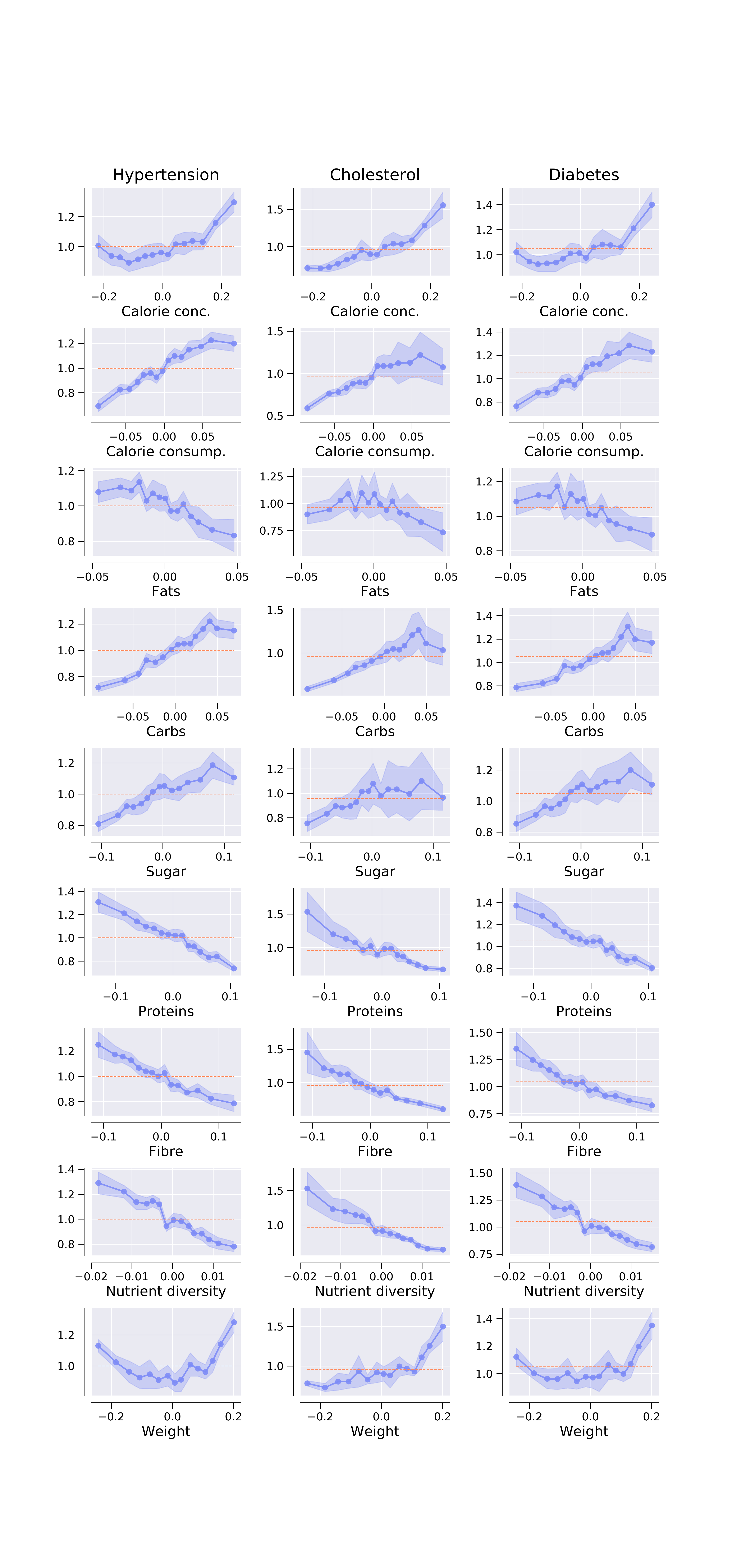}
\caption{The relationship between hypertension, cholesterol, and diabetes (columns) and food-related metrics (rows). On the $x$-axis, we represent the relative food-consumption values compared to the average (set at 0), which are computed with Formula~(\ref{eqn:premium}). On the $y$-axis, we represent the per-capita disease prevalence, which is computed with Formula~(\ref{eqn:diabetes_at_area}): a value of 1 for a disease means that each resident takes, on average, 1 medication for that disease in the year. The dotted red line indicates the average disease prevalence across all areas. The shaded areas show $95\%$ confidence intervals.}
\label{fig:trends}
\end{figure}

We compute the Spearman rank correlation between disease prevalence (as per Formula (\ref{eqn:diabetes_at_area})) and all the food-related metrics (as per Formulae (\ref{eqn:calorie_consumption})-(\ref{eqn:item_weight})). As shown in Figure~\ref{fig:correlations}, calorie consumption is strongly correlated with cholesterol and hypertension,  while calorie concentration is strongly correlated with diabetes (RQs1+2). To check whether the relationships between calories and chronic diseases are linear or not, we produce a set of x-y plots arranged in three columns and nine rows in Figure~\ref{fig:trends}. Each column corresponds to one of the three chronic diseases, and each row corresponds to one of the features derived from our food purchases. For example, in the first row, we have plots that relate hypertension, cholesterol, and diabetes to calorie concentration; in the second row, instead, we have plots that relate these three diseases to calorie consumption. In both cases, we see that, as calories increase, the prevalence of any of the three diseases increases, as one would expect. To ease interpretation of these $x$-$y$ plots, we rescaled the $x$-axis. For example, we normalize the average item's weight in area $a$ as follows:
\begin{equation}
relative\text{-}item\text{-}weight@a = \frac{item\text{-}weight@a-\mu(item\text{-}weight)}{\mu(item\text{-}weight)},
\label{eqn:premium}
\end{equation}
where $\mu(item\text{-}weight)$ is the average weight across all areas. If the rescaled value is zero, then the area's weight is equal to the average value in London. If the value is 0.1, then the area's weight is 10\% higher than the average value. By observing, for example, the resulting plot related to calorie concentration (first row in Figure~\ref{fig:trends}), we see that, as the consumption exceeds the average value (i.e., $x>0$), the per-capita prevalence of any of the three diseases considerably increases.

We also see that the four main nutrients are associated with the three diseases in expected ways: carbohydrates, fat and sugar are positively associated with the three diseases, while fibres are negatively associated (RQs3+5) (Figure~\ref{fig:correlations}). Indeed, the prevalence of each of the diseases increases as carbohydrates (fourth row in Figure~\ref{fig:trends}) increases. The relationships between the diseases and fat (third row in Figure~\ref{fig:trends}) is not as clear cut as one would have expected though: the relationships with fat and saturated fat come with high variability (RQ4). By contrast, more proteins (sixth row) and fibres (seventh row) are associated with lower disease prevalence. 

\begin{figure}[t!]
\centering
\includegraphics[width=0.45\columnwidth]{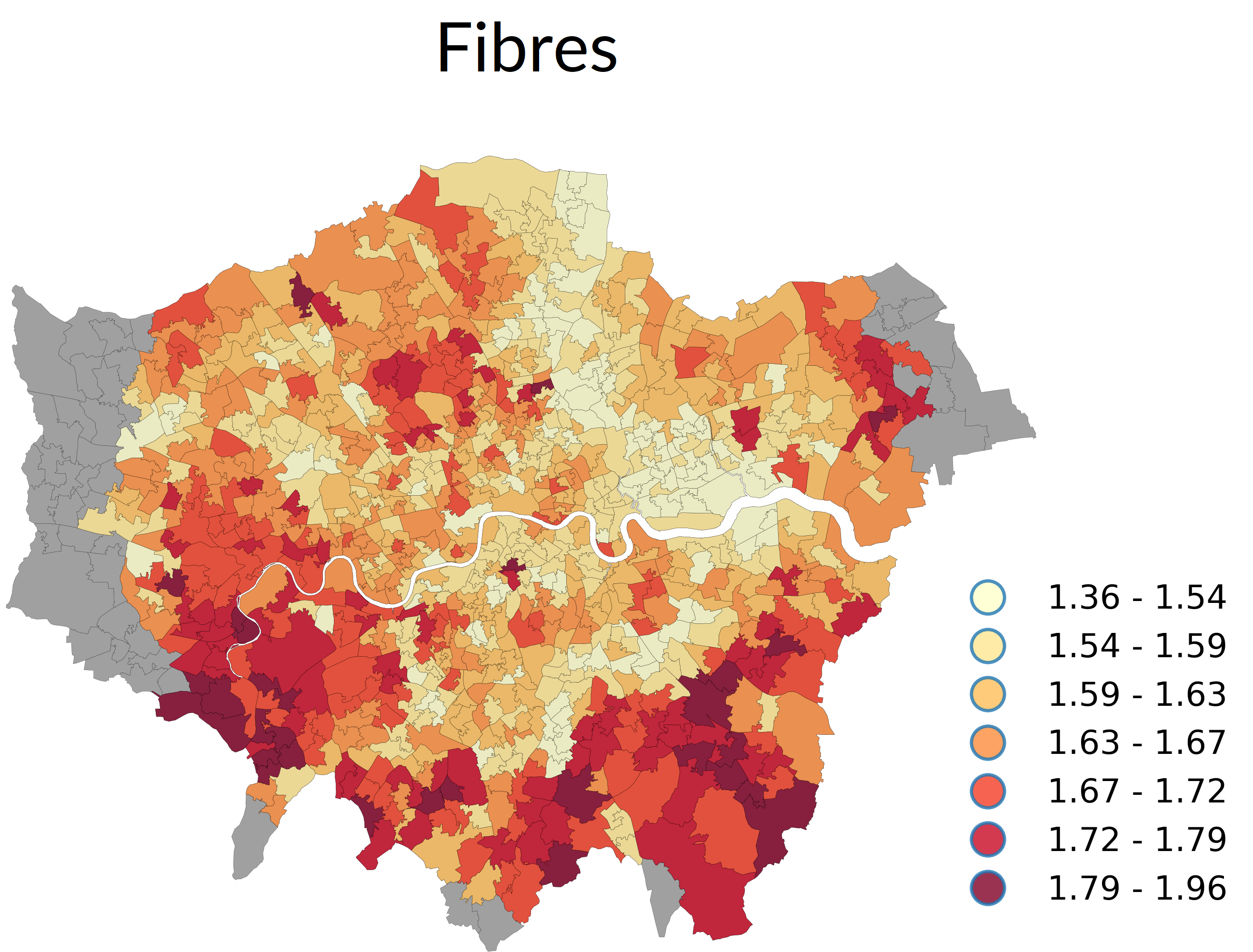}
\includegraphics[width=0.45\columnwidth]{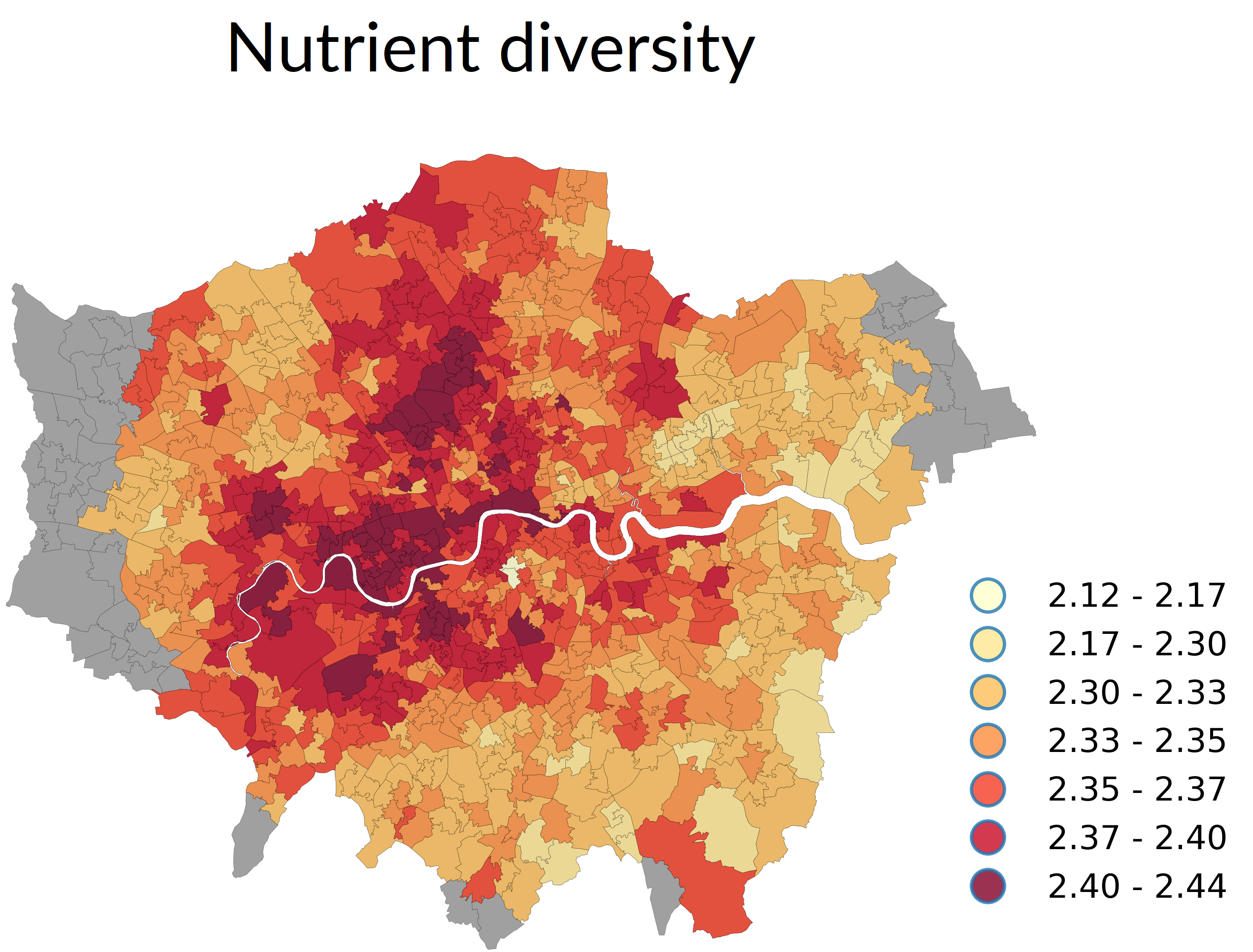}
\caption{London maps reflecting the fraction of an area's total calories coming from fibres as per Formula~(\ref{eqn:nutrients_fraction}) (left), and nutrient diversity as per Formula~(\ref{eqn:nutrients_diversity}) (right).}
\label{fig:nutrient_diversity_map}
\end{figure}

\begin{figure}[t!]
\centering
\includegraphics[width=0.45\columnwidth]{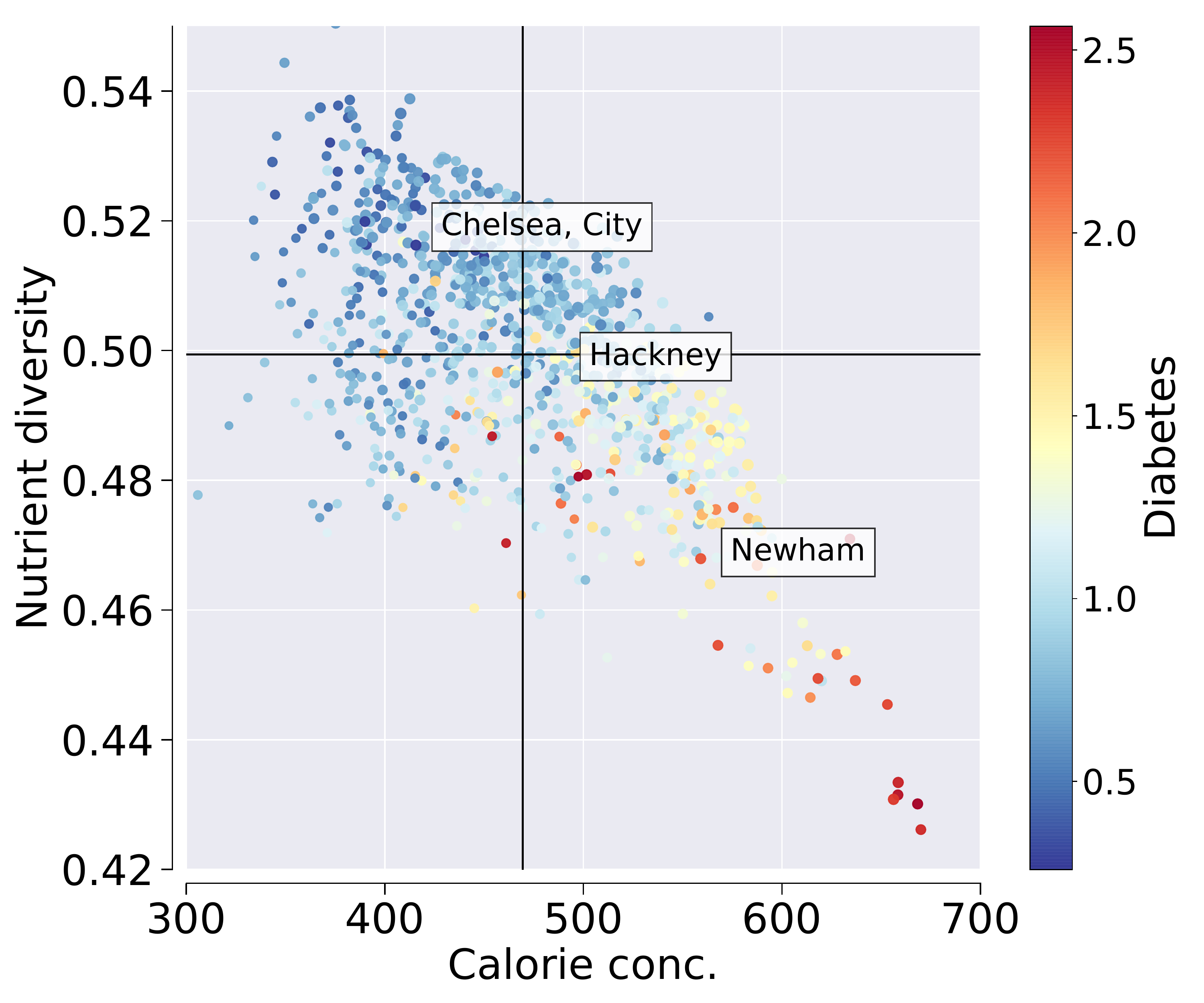}
\includegraphics[width=0.45\columnwidth]{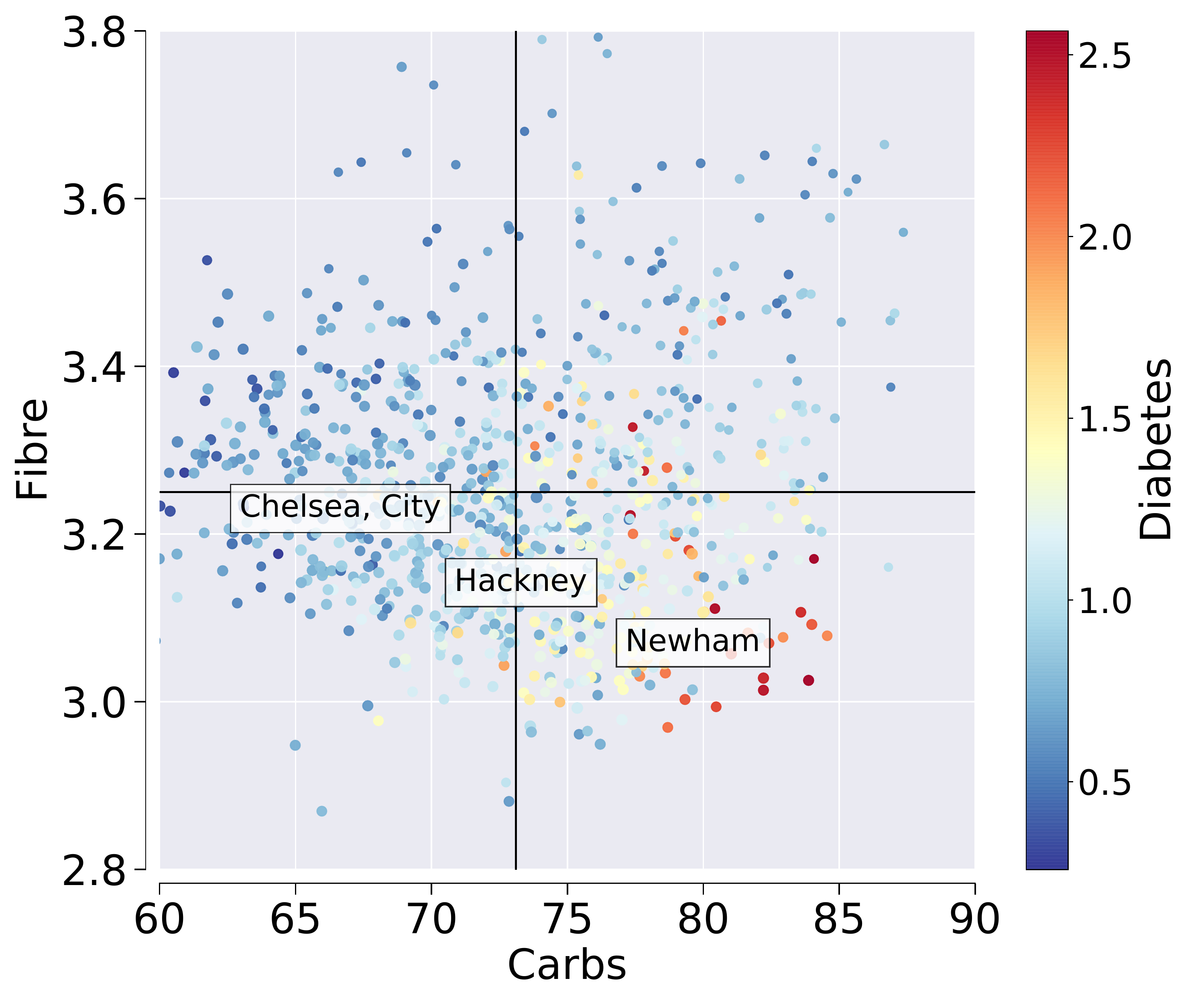}
\caption{Two quadrants that place areas (MSOAs) according to their values for food-related predictor pairs  (computed with Formulae (\ref{eqn:calorie_concentration}), (\ref{eqn:nutrient}), and (\ref{eqn:nutrients_diversity})),  and that colors them according to per-capita prevalence of diabetes (computed with Formula (\ref{eqn:diabetes_at_area})). The horizontal and vertical black lines represent the median values. Healthy areas are at the top-left quadrant, while unhealthy ones are at the bottom-right quadrant.}
\label{fig:quadrants}
\end{figure}

To go beyond individual nutrients, we also find that both nutrient diversity  (mapped in Figure~\ref{fig:nutrient_diversity_map}) and average item weight show high correlations, which of course have opposite signs: item weight (which correlates with  calorie consumption) is positively correlated with disease prevalence, while nutrient diversity is negatively correlated (RQs6+7). Indeed, in Figure~\ref{fig:trends}, one sees that the prevalence of any of the three diseases rapidly decreases with nutrient diversity, while it considerably increases with calorie concentration and average item weight. This is further confirmed by the quadrant in Figure~\ref{fig:quadrants}, which places areas according to the prevalence of pairs of nutrients (computed with Formulae (\ref{eqn:calorie_concentration}), (\ref{eqn:nutrient}), and (\ref{eqn:nutrients_diversity})),  and colors them according to the prevalence of diabetes (computed with Formula (\ref{eqn:diabetes_at_area})): residents of the  City (central London) and Chelsea (West London), who tend to be highly educated and well-off,  consume fibres and diversify nutrients; those of Newham, which is a deprived yet rapidly developing neighborhood in East London, consume considerable quantities of calories, do not diversify their nutrients, and end up with a high prevalence of diabetes; interestingly, the residents of Hackney, which is a deprived yet highly-educated neighborhood in East London, enjoy healthier eating habits (i.e., low consumption of carbohydrates and high nutrient diversity) and do not suffer from diabetes as much as Newham's residents do.

\begin{figure}[t!]
\centering
\includegraphics[width=0.60\columnwidth]{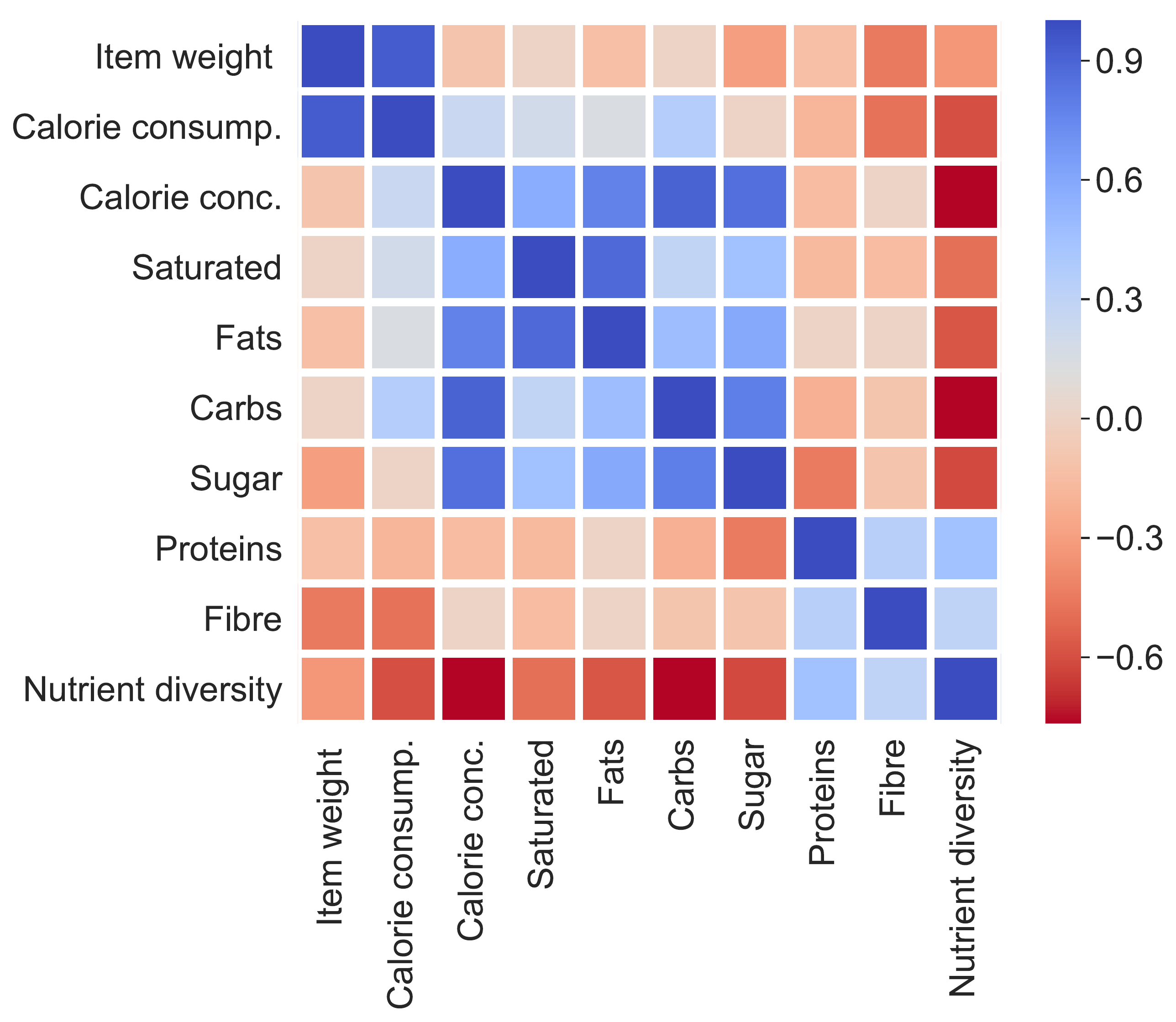}
\caption{Cross correlations among food-related predictors (computed as per Formulae~(\ref{eqn:calorie_consumption})-(\ref{eqn:item_weight})).}
\label{fig:crosscorr}
\end{figure}

So far we have considered our food-related metrics individually. However, these metrics are not orthogonal, and the presence of one is generally associated with the presence of another. In fact, by correlating the presence of a nutrient with the presence of another (cross-correlation matrix in Figure~\ref{fig:crosscorr}), we see that an item's average weight (on the first row of the correlation matrix) is generally not related to any nutrient, as one would expect; carbohydrates (second row) are associated with calorie concentration and sugar (sugar is indeed one type of carbohydrate);  high calorie concentration (third row), in turn, comes with food high in carbohydrates, fat, and sugar; and nutrient diversity (last row) is generally found in food high in proteins and fibres. 


\subsection{Predicting medical prescriptions from nutrients} \label{sec:results:prediction}
As a next step, we go beyond studying correlations and aim at predicting the number of prescriptions from the food data. To do that, we first need to account for the dependencies between nutrients. Also, we should account for any factor other than nutrients that impacts health outcomes. The literature typically controls for socio-economic conditions, which have been shown to be a proxy for access to knowledge and capabilities, including access to nutritional information and physical exercising~\cite{gidlow2006systematic}. To account for all these aspects, we use linear regression analysis. The prevalence of each of the three chronic diseases is the outcome variable of an ordinary least squares regression, while our food-related metrics and a set of control variables are the predictor variables. Where necessary, predictor variables undergo a logarithmic transformation, and in addition, we apply a min-max rescaling of each variable in the range $[0,1]$, which allows us to assess the relative influence of each factor: the larger the absolute value of the coefficient associated to a feature, the higher the relative importance of feature in predicting the outcome.

\begin{table}[ht!]
\caption{Linear regressions that predict the three chronic diseases from individual nutrients and socio-demographic control variables (income, gender, age, education level).}
\label{tab:regression_nutrients}
\begin{tabular}{cccc}
\specialrule{.1em}{.05em}{.05em} 
\multicolumn{4}{c}{\textbf{Hypertension}}\\
\textbf{Feature}	& \textbf{Coefficient} &	\textbf{Std. error} &	\textbf{p-value}  \\
\hline
$\alpha$ (intercept) &	0.3241 & 0.050 & $<$0.001 \\
Carbs	& 0.7652	& 0.075	& $<$0.001 \\
Fats & 0.3225	& 0.066	& $<$0.001 \\
Sugar	& 0.2402	& 0.072	& 0.001 \\
Proteins	& -0.2868	& 0.050	& $<$0.001 \\
Fibre	& -0.0627	& 0.051	& 0.216 \\
Income 	& 0.0477	& 0.042	& 0.259 \\
\%Females	& -0.2124& 	0.038& 	$<$0.001 \\
Average age	& 0.1664	& 0.037	& $<$0.001 \\
Education	& 0.2451	& 0.042	& $<$0.001 \\
\hline
\multicolumn{2}{l}{Durbin-Watson stat. = 2.048} 	& \multicolumn{2}{l}{\textbf{Adj $R^2$ = 0.388}}   \\
& & & \\
\specialrule{.1em}{.05em}{.05em} 
\multicolumn{4}{c}{\textbf{Cholesterol}}\\
\textbf{Feature}	& \textbf{Coefficient}	& \textbf{Std. error}	& \textbf{p-value} \\
\hline
$\alpha$ (intercept)	& 0.2645	& 0.047	& $<$0.001 \\
Carbs	& 0.5877	& 0.070	& $<$0.001 \\
Fats	& 0.3382	& 0.062	& $<$0.001 \\
Sugar & 0.2441	& 0.067	& $<$0.001 \\
Proteins	& -0.2745	& 0.046	& $<$0.001 \\
Fibre	& -0.0268	& 0.047	& 0.569 \\
Income 	& -0.0184	& 0.039	& 0.640 \\
\%Females	& -0.2322	& 0.036	& $<$0.001 \\
Average age	& 0.1272	& 0.035	& $<$0.001 \\
Education	& 0.1751	& 0.039	& $<$0.001 \\
\hline
\multicolumn{2}{l}{Durbin-Watson stat. = 2.001} 	& \multicolumn{2}{l}{\textbf{Adj $R^2$ = 0.345}}   \\
& & & \\
\specialrule{.1em}{.05em}{.05em} 
\multicolumn{4}{c}{\textbf{Diabetes}}\\
\textbf{Feature}	& \textbf{Coefficient}	& \textbf{Std. error}	& \textbf{p-value} \\
\hline
$\alpha$ (intercept)	& 0.5073	& 0.041	& $<$0.001 \\
Carbs	& 0.57659	& 0.061	& $<$0.001 \\
Fats	& 0.5002	& 0.054	& $<$0.001 \\
Sugar	& 0.4992	& 0.059	& $<$0.001 \\
Proteins	& -0.5137	& 0.041	& $<$0.001 \\
Fibre	& -0.1312	& 0.041	& 0.002 \\
Income 	& -0.1222	& 0.034	& $<$0.001 \\
\%Females	& -0.3536	& 0.031	& $<$0.001 \\
Average age	& -0.0290	& 0.030	& 0.342 \\
Education	& -0.0947	& 0.035	& $<$0.006 \\
\hline
\multicolumn{2}{l}{Durbin-Watson stat. = 2.000} 	& \multicolumn{2}{l}{\textbf{Adj $R^2$ = 0.598}} \\
\end{tabular}
\end{table}

\begin{table}[ht!]
\caption{Linear regressions that predict the three chronic diseases from item weight and nutrient diversity (plus control variables such as income, gender, age, education level).}
\label{tab:regression_diversity}
\begin{tabular}{cccc}
\specialrule{.1em}{.05em}{.05em} 
\multicolumn{4}{c}{\textbf{Hypertension}}\\
\textbf{Feature}	& \textbf{Coefficient} &	\textbf{Std. error} &	\textbf{p-value}  \\
\hline
$\alpha$ (intercept)	& 0.5582	& 0.064	& $<$0.001 \\
Calorie consumption& 	0.30228	& 0.070	& $<$0.001 \\
Nutrient diversity	& -0.5182	& 0.069	& $<$0.001 \\
Income 	& 0.0615	& 0.041	& 0.131 \\
\%Females	& -0.2210	& 0.038	& $<$0.001 \\
Average age	& 0.1627	& 0.037	& $<$0.001 \\
Education	& -0.2309& 	0.041	& $<$0.001 \\
\hline
\multicolumn{2}{l}{Durbin-Watson stat. = 2.033	} 	& \multicolumn{2}{l}{\textbf{Adj $R^2$ = 0.377	}}   \\
& & & \\
\specialrule{.1em}{.05em}{.05em} 
\multicolumn{4}{c}{\textbf{Cholesterol}}\\
\textbf{Feature}	& \textbf{Coefficient}	& \textbf{Std. error}	& \textbf{p-value} \\
\hline
$\alpha$ (intercept)	& 0.5465	& 0.059	& $<$0.001 \\
Calorie consumption	& 0.1395	& 0.064	& 0.03 \\
Nutrient diversity	& -0.4943	& 0.064	& $<$0.001 \\
Income 	& 0.0017	& 0.037	& 0.96 \\
\%Females	& -0.2364	& 0.035	& $<$0.001 \\
Average age	& 0.11790	& 0.034	& 0.001 \\
Education	& -0.1785	& 0.037	& $<$0.001 \\
\hline
\multicolumn{2}{l}{Durbin-Watson stat. = 1.981} 	& \multicolumn{2}{l}{\textbf{Adj $R^2$ = 0.344	}}   \\
& & & \\
\specialrule{.1em}{.05em}{.05em} 
\multicolumn{4}{c}{\textbf{Diabetes}}\\
\textbf{Feature}	& \textbf{Coefficient}	& \textbf{Std. error}	& \textbf{p-value} \\
\hline
$\alpha$ (intercept)	& 0.7582	& 0.038	& $<$0.001 \\
Calorie concentration	& 0.1301	& 0.028	& $<$0.001 \\
Nutrient diversity	& -0.6353	& 0.043	& $<$0.001 \\
Income 	& -0.0790	& 0.034	& 0.019 \\
\%Females	& -0.3693	& 0.031	& $<$0.001 \\
Average age	& -0.0606	& 0.030	& 0.042 \\
Education	& 0.1047	& 0.032	& 0.001 \\
\hline
\multicolumn{2}{l}{Durbin-Watson stat. = 1.964} 	& \multicolumn{2}{l}{\textbf{Adj $R^2$ = 0.585}} \\

\end{tabular}
\end{table}

We first try a regression that considers individual nutrients (carbohydrates, fats, sugar, proteins, fibres) as independent variables. Table 1 shows the results and suggests that carbohydrates, fats and sugar are associated with the three chronic diseases, while the presence of proteins and fibres counters that association. Among the control variables, income has very little predictive power when combined with the other factors (it has always either a low coefficient or a high $p$-value). Education and gender are more informative predictors among the socio-demographic variables. In particular, the prevalence of diabetes seems to be more prevalent among males, which is in line with previous findings~\cite{logue2011men}. Overall, nutrients and demographic features jointly explain more than one third of the variability in the linear regressions for hypertension ($R^2 = 0.388$) and cholesterol ($R^2 = 0.345$), and almost 60\% in the regressions for diabetes ($R^2 = 0.598$), and such an explanatory power is not impacted by any autocorrelation. That is because the Durbin-Watson statistic values reported at the bottom of all the regression tables reflect the impact of autocorrelations on the residuals, are defined in the range [0,4], and, in our case, take values close to 2, which indicate no autocorrelation. 

Previously, we have seen that nutrient diversity and calorie consumption tend to be highly correlated with disease prevalence. As such, one might now wonder whether a linear regression analysis solely based on the combination of nutrient diversity and calorie intake would be informative. Indeed, we find that it is (Table 2), and in two main ways. First, based on the regression coefficients, both indicators seem to matter, with nutrient diversity being the most powerful of the two. Second, after controlling for socio-economic variables, these two metrics alone explain up to 38\% of the variance in the prevalence of hypertension, up to 34\% for cholesterol, and up to 59\% for diabetes.

\begin{figure}[t]
\centering
\includegraphics[width=0.85\columnwidth]{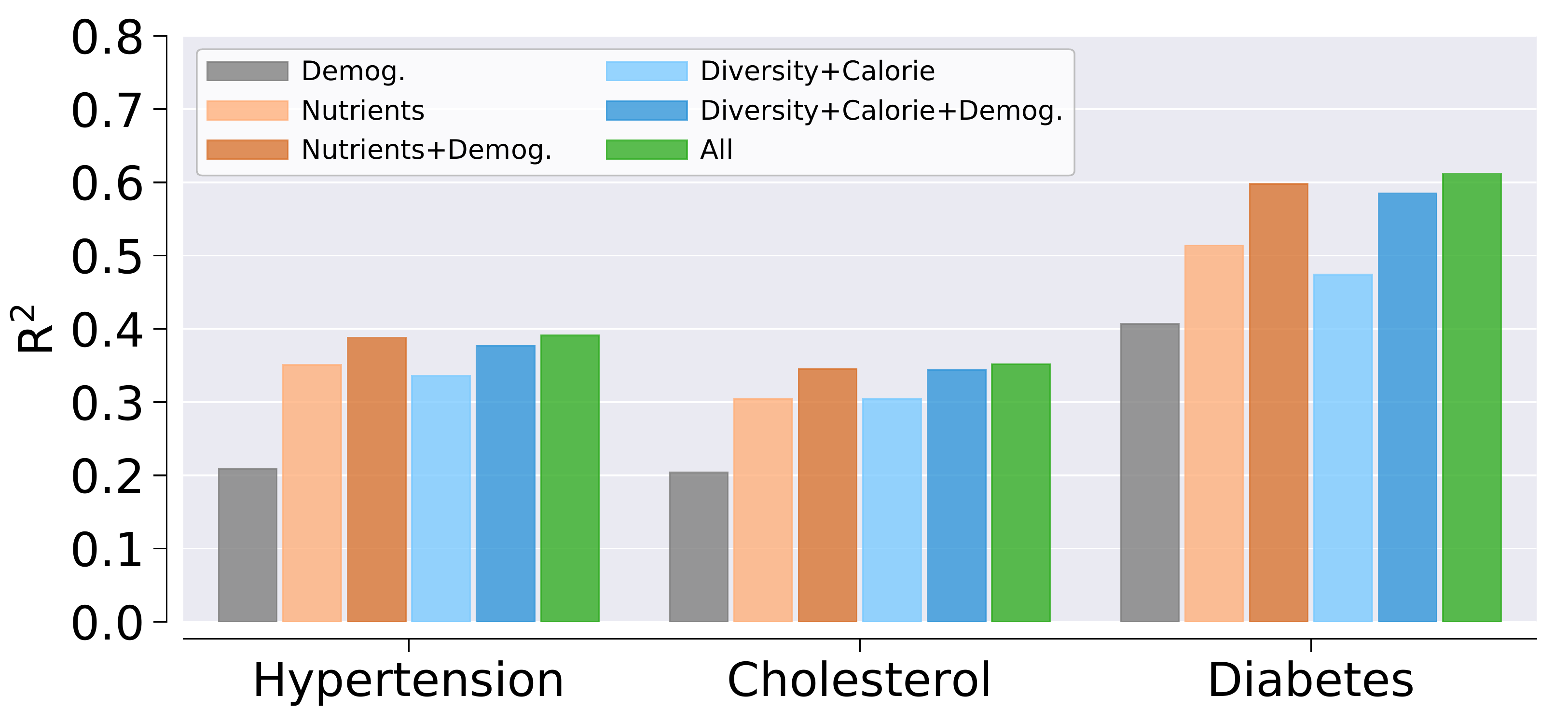}
\caption{$R^2$ values of  regressions having different combinations of features.}
\label{fig:sensitivity}
\end{figure}

To better make sense of the predictive power of our features, we run a sensitivity analysis where we measure the $R^2$ values for regressions run with different feature sets (Figure~\ref{fig:sensitivity}). First, we note that demographic features alone are considerably less predictive than nutrients alone, although they improve the overall accuracy when combined with nutrients. Second, and more interestingly,  the prediction performance of only the combination of nutrient diversity and calorie intake is comparable to the more complex combination of all the individual nutrients. 

\begin{table}[ht!]
\caption{Accuracy of Random Forest classifier in predicting the prevalence of diseases in London areas. The results of two classifiers are reported: i) binary classification of areas in the top and bottom quartiles of the three diseases' prevalence; ii) ternary classification where an equally-sized class containing training instances randomly sampled from the two central quartiles is added. The predictive features are six: gender, average age, education level, item weight, nutrient diversity, and calorie concentration. The accuracy of a random baseline classifier is 0.5 for the binary case, and 0.33 for the ternary case. Numbers in parenthesis represent the standard deviation on the 10-fold cross validation.}
\label{tab:prediction}
\begin{center}
\begin{tabular}{cccccc}
\specialrule{.1em}{.05em}{.05em} 
& & \multicolumn{4}{c}{\textbf{Accuracy}}\\
\cline{3-6}
& \textbf{Medicine}	& \textbf{Random} & \textbf{Demographic}	& \textbf{Diversity + Calorie}	& \textbf{All} \\
\hline
& Hypertension & 0.50 & 0.60 {\tiny(0.06)}	& 0.80 {\tiny(0.05)}	& 0.82 {\tiny(0.05)} \\
Binary & Cholesterol & 0.50	& 0.59 {\tiny(0.06)} 	& 0.81 {\tiny(0.05)}	& 0.81 {\tiny(0.05)} \\
& Diabetes & 0.50	& 0.79 {\tiny(0.06)}	& 0.86 {\tiny(0.05)}	& 0.91 {\tiny(0.04)} \\
\hline
& Hypertension & 0.33	& 0.40 {\tiny(0.05)}	& 0.54 {\tiny(0.05)}	& 0.57 {\tiny(0.04)} \\
Ternary & Cholesterol & 0.33	& 0.41 {\tiny(0.05)}	& 0.53 {\tiny(0.06)}	& 0.54 {\tiny(0.07)} \\
& Diabetes & 0.33	& 0.53 {\tiny(0.04)} & 0.63 {\tiny(0.04)}	& 0.68 {\tiny(0.03)} \\
\end{tabular}
\end{center}
\end{table}

Finally, we build classification models that can identify areas that are healthy or unhealthy in terms of the prevalence of the three chronic diseases. We first formulate this classification problem as a binary classification: the goal is to identify areas that fall into the top and bottom quartiles of each of the three diseases (higher scores corresponded to higher presence of a disease). As such, we define the response variable $y_i$ as 0, if area $i$ is in the first quartile of the disease distribution; and as 1, if it is in the last quartile. This formulation effectively prunes the middle quartiles and makes it possible to focus on the classification of extreme samples. In so doing, we also ensure a roughly 1:1 ratio of positive to negative examples in each class. In a second experiment, we define a 3-class classification problem by keeping the points in the two previous classes (top and bottom quartiles) and defining a third class. This class has the same number of points as the two other classes, and these points are randomly selected from the mid quartiles. We compute the mean accuracy of 10 iterations using a Random Forest classifier in both experiments. Results are reported in Table 3. The performance of each model can be interpreted relative to a baseline random classifier, which after a sufficient number of iterations averages out with an accuracy of 0.5 for the binary case, and 0.33 for the ternary case. We then test three combinations of predictors, one at a time: the demographic features (gender, age, income, education), the two most predictive features from the food data (i.e., calorie concentration and nutrient diversity), and those two combinations jointly. As expected from the previous analysis, demographic factors have the lowest predictive power yet are orthogonal to  food-related predictors.  The binary classifier that uses all features correctly identifies (un)healthy areas 91\% of the times for diabetes, 82\% of the times for hypertension, and 81\% of the times for cholesterol.

\section{Discussion} \label{sec:discussion}

\subsection{Main results} \label{sec:discussion:results}

All the previous results suggest that, in London, socio-economic conditions matter far less than what people eat. As opposed to having high levels of education or of median income, eating less calories and opting for a diet with diverse nutrients are both strongly associated with healthy areas. Indeed, one of the surprising results from the regression analysis is that income is not a significant predictor. Previous studies have shown that income is correlated with general health conditions including mental disorders, self-reported bad health, and lower chances of long-term survival~\cite{graham2004socioeconomic}. Yet not all diseases equally interact with socio-economic variables. In contrast to the prevalence of cancer, the prevalence of  illnesses connected to the metabolic syndrome such as circulatory diseases or high cholesterol  does not significantly change across socio-economic groups~\cite{cookson2016socio}. Indeed, previous  population surveys point out that, after controlling for education, the link between metabolic syndrome and other socio-economic factors such as racial background~\cite{williams1997racial} or income~\cite{loucks2007socioeconomic} weakens markedly, which is in line with our results.

In terms of which nutrients matter the most, as opposed to unhealthy areas, healthy ones tend to buy more fibres and far less carbohydrates (including sugar). Also, it is less about calorie consumption and  more about calorie concentration, which have been previously found to lead to forms of addiction~\cite{Volkow2012}.  By combining all the predictors together, one obtains a model that is not only \textit{descriptive} of how health outcomes are associated with food purchases but also \textit{predictive} of such outcomes: it turns out that, from food purchases, we can accurately predict whether an urban area will suffer from diabetes, for example.

\subsection{Theoretical implications}\label{sec:discussion:theoretical}

This study has two main theoretical implications. The first is a call for enlightened nutrition research. The question for food companies is how to continue to make money even as they cut calories. The answer might come from a shift in how food companies have approached the formulation of their products so far.  Food product research has focused its attention on taste, not nutrition. That needs to change. The combination of nutritional research with recent advances in biomedical research promises to create foods that are not only delicious but also provide concrete medicinal benefits. 

The second theoretical implication has to do with studying how entire neighborhoods eat. Past research has explored two relationships. The first is between ``where people live'' and their health: income inequality, unemployment rates, and education have all been shown to relate to people's health. The second is between ``what people eat'' and their health: nutrition research has long tested associations between eating patterns and health outcomes with survey data. A third relationship transitively follows which has never been tested before: the relationships between ``where people live'' and ``what they eat''. We have now tested it and found that, indeed, healthy neighborhoods eat less and diversify nutrients more than what neighborhoods suffering from chronic diseases tend to do. 

\subsection{Practical implications}\label{sec:discussion:practical}

This study suggests practical implications for a variety of stakeholders. We have shown that unhealthy products have a negative impact on community health. The bad news is that many unhealthy products are very popular. The good news is that as many as five stakeholders have incentives for change: food companies do not wish to be seen as the cause of people's obesity; insurance companies (especially those in life insurance) have our health at heart; technology companies are entering the digital health market; governments want to be seen to act; and local communities increasingly want to be empowered to tackle their own needs.

\vspace{4pt}\textbf{Food Companies}. By simply cutting out bad ingredients, adding good ones or introducing new products, the food industry could reformulate their offering and elaborate plans to improve nutrition.

\vspace{4pt}\textbf{Insurance Companies}. There is at least one corporate sector that benefits from keeping people healthy: insurance companies.  Our study encourages new partnerships between insurance firms and large grocery retailers on, for example, data sharing initiatives. Also, retailers could make anonymized purchased data publicly available and launch ``hackathons''. These are meetings in which participants are asked to come up with a solution to a problem within a day or two, and some of the teams generally offer effective solutions at little cost (the winning team is typically awarded a prize). 

\vspace{4pt}\textbf{Technology Companies}. Predictive analytics and wearable sensors will transform how people manage their health. A smartphone app might be able to warn users that, based on which foods they share on social media and what their wearable sensors measure, they will exacerbate a heart condition. The app could even suggest which foods to eat---foods that are both pleasurable and nutritious.

\vspace{4pt}\textbf{Public authorities}. In the past, governments have focused on treating diseases rather than preventing them. Yet state-based prevention strategies might be justified, not least because of externalities. Unhealthy eating harms not only oneself but also others, in that, it results in additional costs for health care. Public authorities could intervene in three main ways. 
\begin{itemize}
	\item \textit{Taxing}. Taxing fat and subsidizing healthy eating is one way of tackling the obesity problem. However, a recent study showed that taxing fat might not help~\cite{muller16}. In the study, subsidies were given to encourage all income groups to buy more fruit and vegetables. Women on higher incomes bought more fruit and vegetables than usual, while those on lower incomes changed their habits less. As a result, women on lower incomes paid much more for food (as taxes were on the food they ate most), and the inequality between the two groups widened. Taxes and subsidies might not change people's habits, and other strategies are needed---notably education.
	\item \textit{Educating}. One simple state intervention is the launch of new educational programs that inform people about the dangers of not eating well. It has been shown that a short-lived change in diet have long-term consequences. A three-week change of diet aimed at reducing cravings for salt, sugar, and fat has been shown to change participants' taste buds~\cite{grieve03}. 
	\item \textit{Nudging}. A more viable alternative would be to nudge citizens into healthy behavior. The idea is to provide small impulses so that healthy becomes the obvious choice.
\end{itemize}

\vspace{4pt}\textbf{Local Communities}. We have shown that, by mining publicly available prescription data, we are able to identify (un)healthy areas. Mining digital health data reflects concrete health outcomes and might well benefit local communities by enabling residents to hold local authorities to account. Additionally, a city health monitor could help assess the benefits of implementing different policies.

\subsection{Limitations}\label{sec:discussion:limitations}

\noindent\textbf{Sample bias.} Our data  comes from one grocery retailer and concerns only those customers who have opted for the use of a loyalty card. Furthermore, the data is anonymized and does not contain personal information such as age or gender. Future work should use additional geographic data to quantify sample biases.

\vspace{4pt}\noindent\textbf{Limited explanatory power.} Our study does not fully explain health outcomes, and rightly so. Our food data does fully cover Londoners' food consumption, and our study does not include any data on another important predictor of health outcomes: physical activity.

\vspace{4pt}\noindent\textbf{Average product.} From our data, we cannot reconstruct the dietary habits of individual customers and, as such, our results reflect the dietary habits of an area in terms of the area's ``average product''.

\vspace{4pt}\noindent\textbf{Causality.} Our results do not speak to causality. Though the causal direction is difficult to determine from observational data, one could consider different temporal snapshots of both sets of data (food purchases and medical prescriptions), and perform a cross-lag analysis.

\subsection{Conclusion}\label{sec:discussion:conclusions}

It was healthy and adaptive for our primate brains to drive us to eat carbohydrates and sugar when only wild grass was at hand. However, carbohydrates (including sugar) are now readily available at every corner. People living in areas of London with higher prevalence of medical conditions linked to the metabolic syndrome seem to surrender to their human instincts and end up buying carbohydrates and sugar to a considerable extent. By contrast, people living in healthy neighborhoods seem to counter their evolutionary adaptation and buy considerable quantities of fibre. This difference in purchases is not explained by socio-economic conditions: income does not matter as much as one expects. By transcending conventional class boundaries, human biases, instead, seem to be the main obstacle to healthy eating. Our study suggests that the ``trick'' to not being associated with chronic diseases is eating less what we instinctively like (by not listening to the dopamine rushes in our brains), balancing all the nutrients, and avoiding the (big) quantities that are readily available. 

In the future, we will explore the impact of two additional factors on health outcomes. The first is the city itself: certain city's forms are more appealing to pedestrians than others and, as such, one might wonder which forms are ``healthier''. The second factor is exercising. We are exploring the possibility of capturing exercising levels across an entire city with wearable devices. Too see why this is important, consider that, by exercising (even a little), an individual boosts his/her immune system, achieves a 20-50 percent reduction in sick days in the short term, and reduces the risk of chronic diseases in the long term~\cite{nieman11}. 

In our cities, food is cheap and exercise discretionary, and health takes its toll. Technology could change that. With modern data analytics, the availability of new open data, recent advances in persuasive computing, and ever increasingly miniaturized health wearables, modern technologies are now best positioned to help people counter the dopamine rushes coming from sugar and fat, eat better, and exercise more.

\begin{backmatter}

\section*{Abbreviations}

\textbf{GP}: General Practitioner

\textbf{ONS}: Office for National Statistics

\textbf{OA}: Output Areas

\textbf{LSOA}: Lower Super Output Areas

\textbf{MSOA}: Medium Super Output Areas

\textbf{LA}: Local Authority

\textbf{IMD}: Index of Multiple Deprivation

\section*{Availability of data and material}
Our food dataset is available upon request. To preserve customer privacy, the data has been aggregated at MSOA level. After publication, the dataset will be uploaded to a third party repository with a permanent URI and will be also available on the project's site \url{http://goodcitylife.org/food}.

\section*{Funding}
Not applicable

\section*{Competing interests}
 The authors declare that they have no competing interests. This work was done while LMA and DQ were employees of Nokia Bell Labs and LDP was employee of Tesco Labs. The authors' employers provided support in the form of salaries but did not have any additional role in the study design, data collection and analysis, or preparation of the manuscript. All work was done as part of the respective authors' research, with no additional or external funding. 

\section*{Author's contributions}
LMA and RS worked on the prescription data collection and performed the analysis. DQ set the hypotheses and research questions. LDP worked on the food data collection. All authors contributed to establish the methodology and to write the manuscript.

\bibliographystyle{bmc-mathphys} 


\end{backmatter}
\end{document}